# Computational Reverse-Engineering Analysis for Scattering Experiments for Form Factor and Structure Factor Determination ('P(q) and S(q) CREASE')[†]

Christian M. Heil,[1] Yingzhen Ma,[2] Bhuvnesh Bharti,[2] and Arthi Jayaraman[1,3,*]

[1]Department of Chemical and Biomolecular Engineering, University of Delaware, 150 Academy St., Newark, DE 19716. USA.

[2]Cain Department of Chemical Engineering, Louisiana State University, 3307 Patrick F. Taylor Hall, Baton Rouge, LA 70803. USA

[3]Department of Materials Science and Engineering, University of Delaware, 201 DuPont Hall, Newark, DE 19716. USA.

*Corresponding author arthij@udel.edu

[†]Electronic supplementary information (ESI) available.

**Abstract**

In this paper we present an open-source machine learning (ML) accelerated computational method to analyze small angle scattering profiles [I(q) vs. q] from concentrated macromolecular solutions to simultaneously obtain the form factor P(q) (*e.g.*, dimensions of a micelle) and structure factor S(q) (*e.g.*, spatial arrangement of the micelles) without relying on analytical models. This method builds on our recent work on Computational Reverse Engineering Analysis for Scattering Experiments (CREASE) that has either been applied to obtain P(q) from dilute macromolecular solutions (where S(q) ~1) or to obtain S(q) from concentrated particle solution when the P(q) is known (*e.g.*, sphere form factor). This paper's newly developed CREASE that calculates P(q) and S(q), termed as 'P(q) and S(q) CREASE' is validated by taking as input I(q) vs. q from *in silico* structures of known polydisperse core(A)-shell(B) micelles in solutions at varying concentrations and micelle-micelle aggregation. We demonstrate how 'P(q) and S(q) CREASE' performs if given two or three of the relevant scattering profiles - $I_{total}(q)$, $I_A(q)$, and $I_B(q)$ - as inputs; this demonstration is meant to guide experimentalists who may choose to do small-angle X-ray scattering (for total scattering from the micelles) and/or small-angle neutron scattering with appropriate contrast matching to get scattering solely from one or the other component (A or B). After validation of 'P(q) and S(q) CREASE' on *in silico* structures, we present our results analyzing small angle neutron scattering profiles from a solution of core-shell type surfactant coated nanoparticles with varying extents of aggregation.

**Table of Contents Figure**

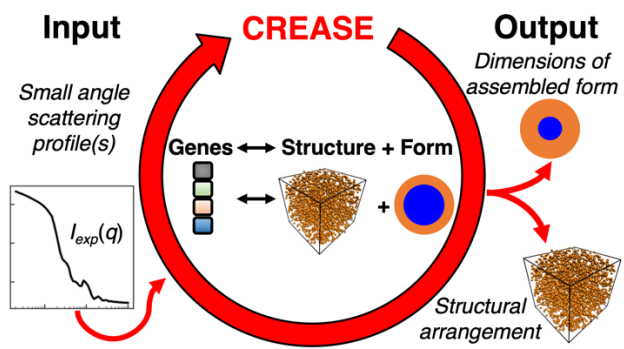

# I. Introduction

Multiscale hierarchical structures achieved through self-assembly or directed assembly of nanoparticles and/or macromolecules in solutions/melts lead to materials applicable in medicine, sensing, separations, optics, photonics, and electronics.[1-12] The features of the assembled structure depend on materials design (*e.g.*, polymer chemistry, sequence, and architecture or nanoparticle size, shape, and surface functionalization) and processing conditions (*e.g.*, solution temperature and solvent quality).[1, 3, 4, 13-16] In all cases, the characterization of the structure at different scales (*i.e.*, domain shape and size, chain size, spatial arrangement) is a necessary step to optimizing and engineering useful materials.

The characterization of assembled macromolecular structures is typically performed using microscopy (*e.g.*, transmission electron microscopy (TEM), scanning electron microscopy (SEM), cryogenic TEM (cryo-TEM), and atomic force microscopy (AFM))[17-20] and small angle neutron or X-ray scattering (SANS or SAXS, respectively).[21-25] Microscopy characterization methods produce images that readily allow for direct identification of the shape(s) and dimensions. However, electron microscopy also requires a high degree of sample processing such as drying or freezing, which often leads to artifacts and inhibits the 'true' structure determination. Further, the images are typically only 2D slices of the 3D self-assembled structure, and the image resolution can prevent quantification of small length scales. Such limitations can be readily overcome using *in situ* small angle scattering characterization methods. Small angle scattering techniques can analyze macromolecular self-assembled structures over a broad range of length scales from μm down to Å. Additionally, scattering contrast matching of the solution with different polymer chemistries enables selective scattering contribution (and thereby characterization) of certain regions or parts of the assembled structure.[26] Due to these reasons, SANS and SAXS can be powerful tools for understanding macromolecular assembled structures. Unlike microscopy, however, the SAXS and SANS output is a 2D or 1D scattering intensity (I(q)) versus scattering wavevector (q) which often requires complex steps for correct interpretation of the relevant dimensions and shape of the assembled structures.

For solutions of macromolecules or particles, small angle scattering methods characterize structural information including the macromolecular assembly (*e.g.*, micelle) or particle features (*e.g.*, rod, ellipsoid, sphere) as well as their spatial arrangement (*e.g.*, assembly of the micelles or disordered/ordered aggregate of particles). For such cases, the SANS or SAXS experiments' output I(q) vs. q, is the product of the form factor, P(q), (*i.e.*, particle's size, shape) and the structure factor, S(q), (*i.e.*, spatial arrangement of the particles) (**Figure 1**).[25] During analysis of the I(q) vs. q to obtain such structural information, it is common to use the scattering profile from the system at conditions where the S(q) is ~1 (*e.g.*, dilute macromolecular solution concentration or dispersion of particles at low concentration). This way the scattering profile I(q) vs. q is approximated to be the P(q), and the analysis of the I(q) leads to understanding the form of the assembled macromolecule or isolated particle. This P(q) is then used during analysis of the I(q) vs q for that system at conditions of interest when the structure of the system cannot be ignored (*i.e.*, S(q) is not equal to 1). This type of analysis assumes that form factor P(q) is constant in both conditions, which is valid, for example, for particles that do not change shape or size in the conditions of interest (*e.g.*, with increasing concentration, temperature, salinity). However, such an assumption that the form factor P(q) remains constant, particularly with changing concentration, is not necessarily valid for macromolecular solutions. For example, micelles formed in block copolymer solutions can change size and shape with concentration, making it incorrect to use P(q) identified at low concentration also at high concentration.[27, 28] Furthermore, even in cases where the P(q) does not change with concentration, one would need to do a scattering experiment to determine the P(q) at each condition of interest (solvent conditions, temperature, salt concentration) at low concentration in addition to the desired scattering experiments at the target concentration. Thus, there is a need for an approach that allows for simultaneous identification of P(q) and S(q) and structural interpretation directly at the conditions of interest.

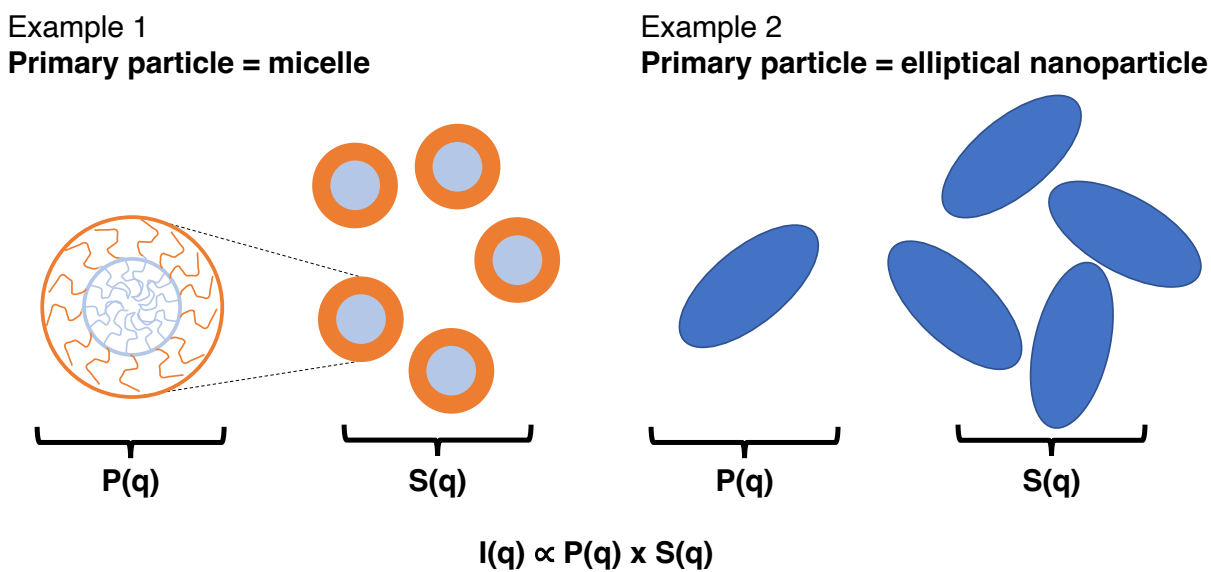

*Figure 1: Two examples of 'primary particle' and schematics of relevant structural information included in the P(q) (i.e., form of 'primary particle') and S(q) (i.e., spatial arrangement of the 'primary particles'). Example 1 depicts the case where the 'primary particle' is a core-shell type micelle formed by amphiphilic macromolecules. In this case one may not be able to assume that the P(q) found at low concentrations (when S(q) ~1) is also applicable at high concentration. The amphiphilic macromolecules may reorganize to form micelles of different dimensions or shapes as the concentration of the solution increases. Example 2 depicts the case where the 'primary particle' is a non-spherical nanoparticle; if the nanoparticle is not expected to evolve in shape*

*and dimension with changing concentration, one can assume that the P(q) found at low concentration is applicable at higher concentration as well.*

In cases where the I(q) ~ P(q) there are many analytical models that have been developed to extract structural information about the 'primary particle' (*i.e.*, macromolecule, macromolecular assembly, coated particle, *etc.*).[29, 30] If the 'primary particle' is a macromolecule then the structural information could be its average radius of gyration and shape of the chain (*e.g.*, cylindrical or globular). If the 'primary particle' is a macromolecular assembly (*e.g.*, micelle) then the structural information would be the core and corona dimensions of the micelle or length and diameter of the cylindrical assembly or core radius and wall thickness of a vesicle. If the 'primary particle' is a bare or coated nanoparticle the structural information could be the shape (*e.g.*, rod, cylinder, sphere, ellipsoid) and relevant dimensions of the nanoparticle and coating. In all cases, the relevant analytical models fit to the I(q) ~ P(q) are developed by making certain assumptions about the 'primary particle.' For example, the analytical P(q) models that one could use to fit to scattering data from canonical spherical micelles in solutions of amphiphilic polymers vary in the assumptions made about the core (*e.g.*, impenetrable core, soft core),[29, 31] corona/shell chain conformation (*e.g.*, semiflexible with exclude volume[32] or Gaussian[33]), and presence or absence of dispersity in the micelle sizes.[33, 34] While the analytical models are available for more conventional chain conformations/assembled structures, due these assumptions the analytical models may not be applicable for structures formed by new polymer chemistries with novel architectures[35] and non-equilibrium structures formed by new assembly techniques.[36, 37] To fill the need for a more generic approach to characterize structure of the 'primary particle' using scattering profiles I(q) ~ P(q) (*i.e.*, conditions where S(q) is ~1), researchers in the research group of Arthi

Jayaraman have developed a computational reverse-engineering analysis for scattering experiments (CREASE) method for a variety of 'primary particles' (micelles,[35, 38, 39] vesicles,[40] and fibrils[41]) bypassing the need for an analytical model.

For systems where the P(q) of the 'primary particle' is known (*e.g.*, nanoparticle with fixed shape and size) but the spatial arrangement of interacting 'primary particles' is unknown, one needs to analyze the S(q) to understand that spatial arrangement. There are many analytical models[42, 43] (*e.g.*, hard sphere[44, 45] and sticky hard sphere[46, 47]) for characterizing the structure of 'primary particles' in a fluid suspension; the fluid suspension assumption, however, does not perform well for systems at high packing fractions (above 0.4) or when there is formation of 'primary particle' aggregates.[48, 49] More complex analytical models for aggregating particles require the user to possess significant *a priori* knowledge of their system as they choose the appropriate analytical structure factor model for the type and quality of structural information extracted (*e.g.*, aggregate radius of gyration and aggregation number).[50] To overcome these limitations with existing analytical S(q) models, we developed a computational method, CREASE, for analyzing the structure of 'primary particles' (*e.g.*, nanoparticle solutions, dense binary nanoparticle mixtures) without requiring a user to select a specific analytical model using substantial *a priori* knowledge from alternative characterization methods.[51, 52] Furthermore, unlike the analytical model fits, this CREASE method provides 3D structural reconstruction of the system being studied which can then be used as an input for other calculations (*e.g.*, resistor network model calculation for electrical conductivity[53] and finite-difference time-domain method for optical properties[54, 55]). However, this recently published CREASE method was designed for interpreting S(q) in mixtures and solutions of nanoparticles where the nanoparticle's (or nanoparticles') P(q) is known *a priori*. For systems where one does not know the P(q) or S(q) *a* *priori*, in this paper we describe an extension of the CREASE methods, 'P(q) and S(q) CREASE' for simultaneously solving for P(q) and S(q). We demonstrate how this approach can also facilitate high-throughput analysis with a fewer number of scattering experiments required to characterize relevant structural features of the system.

In this paper, we validate our new 'P(q) and S(q) CREASE' approach using *in silico* scattering profiles from concentrated solutions of core-shell spherical micelles with differing packing fraction (or concentration), micelle size dispersity, extent of aggregation, and core size to micelle size ratio. We employ a two-step machine learning (ML) approach that significantly reduces analysis time and could be transferred to other related systems (*e.g.*, concentrated solution of interacting vesicles) by simply swapping out the P(q) ML model. Finally, we apply our 'P(q) and S(q) CREASE' method on *in vitro* (*i.e.*, from SANS experiments) scattering profiles of surfactant-coated silica nanoparticles where both the extent of nanoparticle aggregation and surfactant shell thickness characteristics are unknown and anticipated to change with temperature and salt concentration. The 'P(q) and S(q) CREASE' method described in this paper is a broadly applicable computational technique for experimentalists to analyze high-throughput small angle scattering results and obtain reconstructed 3D structural arrangements and form factor of the 'primary particles' in the sample without relying upon approximate analytical models which may be unsuitable for the system under consideration.

**II. Approach**

The presented 'P(q) and S(q) CREASE' method simultaneously solves for the form factor and structure factor in the system and is an augmentation of our previously developed gene-based CREASE method used for structure reconstruction in systems where P(q) is known;[52] we direct the reader particularly to the supplemental information of that paper[52] for all relevant method details and a link to the open-source codes. Briefly, that paper described the development and validation of CREASE for nanoparticle solutions and assemblies. First, we described the structure in the system via a set of 'genes' (*i.e.*, a low dimensional representation of a 3D configuration of particles) which represented information about the material form factor (*i.e.,* nanoparticle shape, size, dispersity) and structure factor (*i.e.,* composition and spatial arrangement). For those nanoparticle systems, the form factor was known *a priori* as the nanoparticle shape, size, and dispersity can be characterized prior to the scattering experiment, and in that case the nanoparticle features were not expected to vary with conditions including salt concentration, temperature, and solvent conditions. Thus, that work focused on systems where the form factor $P(q)$ is known and only the structure factor $S(q)$ had to be solved for. We also direct the reader to other CREASE developments focused on systems where the structure factor is negligible (*i.e.*, $S(q) \sim 1$) to solely identify the form factor of the 'primary particle' (micelles,[35, 38, 39] vesicles,[40] and fibrils[41]). For the work presented in this paper, we focus on systems where the form factor and structure factor are both unknown and non-negligible.

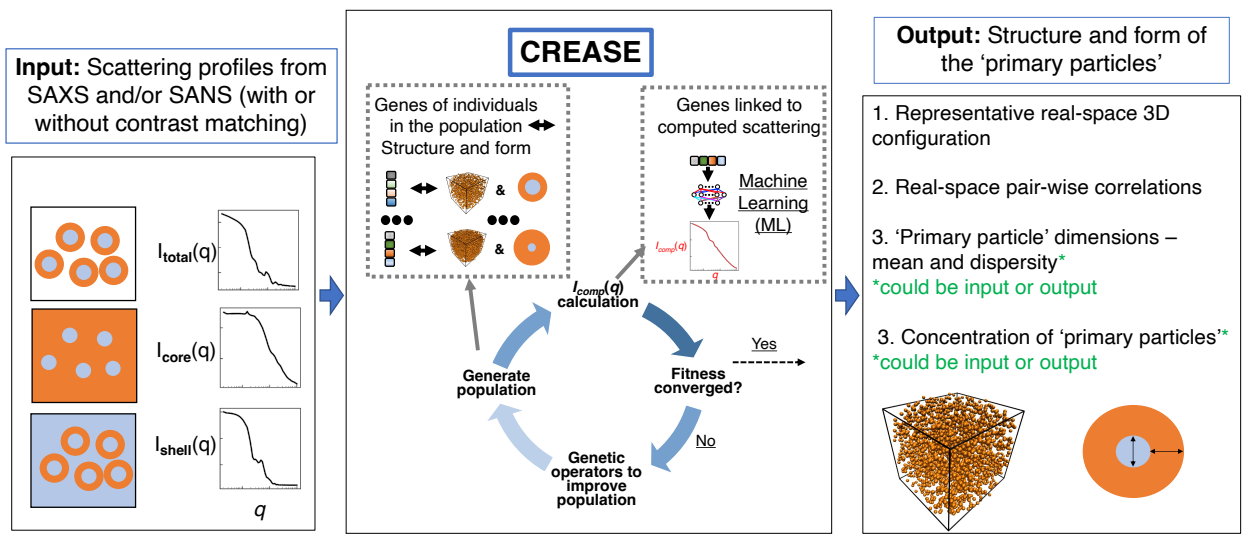

***Figure 2**: Overview of the 'P(q) and S(q) CREASE' workflow. CREASE takes the 'primary particle' solution's small angle scattering profile(s) as input and utilizes a genetic algorithm to optimize for the set of genes (low dimensional feature space corresponding to a structural arrangement, S(q), and primary particle form, P(q)) that possesses a computed scattering profile(s) that most closely matches the input scattering profile(s). CREASE outputs the real-space arrangement of the 'primary particle' as well as the relevant dimensions of the primary particle' (e.g., for spherical core-shell micelles, micelle aggregation or dispersion and micelle's core and shell dimensions). The computed scattering profile(s) in CREASE can be determined directly using the Debye scattering equation or through a machine learning (ML) model.*

**Figure 2** provides a high-level overview of CREASE for simultaneous structure reconstruction and material characterization. CREASE takes the small angle X-ray or neutron scattering profile(s) as an input. A user can input the total scattering of all 'primary particle' components in the system and/or contrast matched scattering profiles to get structure of one component in the 'primary particles' (**Figure 2 input box**). CREASE utilizes a genetic algorithm to optimize for the set of genes (low dimensional feature space corresponding to a structural arrangement, S(q), and form, P(q) of the 'primary particles') that possesses computed scattering profile(s) that most closely matches the input scattering profile(s). The optimization proceeds by starting with a population of individuals each with different sets of genes. The 'fitness' of each individual is determined by comparing its computed scattering profile(s) to the input scattering profile(s). The computed scattering profile(s) in CREASE are determined either by converting the genes to the corresponding structure and using the Debye scattering equation or by a machine learning model trained to link the genes (described in more detail in the next section) to the corresponding scattering profile(s). If the fitness has not converged (*i.e.*, individuals are continuing to improve the match between the computed scattering profile(s) and the input scattering profile(s)), a new generation of individuals is produced by mutation (randomly changing a gene) and/or combination (randomly combining different parts of two individuals' gene sets from the previous generation) to produce a population of new individuals. Once the fitness converges to the optimal set of genes, CREASE outputs the real-space arrangement of the material as well as information on primary particle dimensions (*e.g.*, for the example case of micelle solutions, micelle spatial arrangement, micelle total size, micelle size dispersity, core size, and micelle concentration).

## II. A. Development of the gene set for 'P(q) and S(q) CREASE'

As an example, we describe the set of genes or the feature space to simultaneously solve for the S(q) and P(q) for the cases of concentrated core-corona spherical micelle solution and core-shell nanoparticle solution. The previously developed CREASE for S(q) utilized 10 different genes to reconstruct the structural arrangement of spherical nanoparticles in solution.[52] Genes 1 & 2 related to the average nanoparticle diameter and dispersity; Genes 3 & 4 were the coarse-grained solvent particles to control the degree of aggregation; Gene 5 was the nanoparticle volumetric concentration; Gene 6-8 were how CREASE produced structures with varying degrees of aggregation by controlling the nanoparticle domain size (gene 6), the domain compactness (gene 7), and the spacing between domains (gene 8); gene 9 applied a background scattering intensity; and gene 10 set the systems size by specifying the number of nanoparticles to use in the 3D reconstruction. As we are interested in this paper in analyzing solutions of core-shell or core-corona type 'primary particle', we need to modify and add additional genes relating to the core and shell/corona. Depending on the physics underlying the specific system, the shell might be of constant thickness in all 'primary particles' regardless of core size, the shell with a thickness that scales with the core size, or a shell that has a mean thickness and dispersity independent of the core's mean size and dispersity. Furthermore, one could consider if the material of interest should be treated as a hard core-shell particle or a soft core-shell particle with overlap possible between the shells of multiple 'primary particles'. For each of these cases, we describe how a user could add gene(s) for their specific system.

- For the case of the constant shell thickness, one would need to add an additional gene corresponding to the shell thickness. The genes originally corresponding to the nanoparticle size and dispersity would be adapted to be the core size and dispersity.
- For the case of the shell thickness varying with the core size, one would need to set a gene that relates the shell thickness to the core size. In this work, we consider a linear scaling with size such that the gene represents the ratio of the core radius to total radius. The genes originally corresponding to the nanoparticle size and dispersity would be adapted to be the core size and dispersity.
- For the case of the shell possessing its own average thickness and dispersity, one would need to add two genes corresponding to 1) the average thickness and 2) the thickness dispersity. The genes originally corresponding to the nanoparticle size and dispersity would be adapted to be the core size and dispersity.
- To consider the case of shells potentially allowing overlap, one would add another gene corresponding to the degree of shell overlap allowed in addition to the previous modifications and additions.

## II. B. Conversion of genes to 3D structure and calculation of computed scattering profiles

After the gene(s) have been chosen, a user would then need to adjust how the genes are converted to a structure to ensure the total particle size (core and shell) is properly accounted for and alter how the scattering profile(s) are calculated for the system. For the case of allowing particle overlap, when generating the structure, one would need to reduce the effective total particle size to allow for overlap between 'primary particles.' For the structure generation step of CREASE, the user would reduce the total particle size by the overlap gene's value times the shell thickness (assuming only the shells are soft to allow overlap) or the entire 'primary particle' (assuming the particles can overlap substantially). Once the gene to structure conversion is set, a user would need to specify how the scattering intensity is calculated for each structure. For the scattering intensity calculation, one can use the pairwise Debye scattering equation to directly evaluate the scattering for the structure: [56, 57]

$$I_X(q) = \frac{1}{V_{sample}} \left[ \sum_{i=1}^{N_x} f_i(q)^2 + \sum_{i=1}^{N} \sum_{j \neq i}^{N} f_i(q) f_j(q) \frac{\sin(qr_{ij})}{qr_{ij}} \right] \qquad (1)$$

$I_X(q)$ is the computed scattering intensity (either total scattering, core scattering, or shell scattering) depending on which particle form factor amplitude $f_i(q)$ is used (both the core and shell, only the core, or only the shell). $r_{ij}$ is the pairwise distance between the particle centers, N is the number of particles, and $V_{sample}$ is a scaling term to set $I_X(q)$ to 1.0 at the lowest q value considered to facilitate comparison between input scattering $I_{target}(q)$ and computed scattering $I_{comp}(q)$ profiles. We calculate the f(q) for the total particle, core, and shell by randomly placing point scatterers within the particle:

$$f_X(q) = \sum_{l=1}^{N_p} \sum_{k \neq l}^{N_p} \Delta\rho_l \Delta\rho_k \frac{\sin(qr_{kl})}{qr_{kl}} \qquad (2)$$

$r_{kl}$ is the pairwise distance between the point scatterers' centers, $N_p$ is the number of point scatterers considered, and $\Delta\rho_i$ as the scattering length density (SLD) or electron density difference between the solvent and the component (core or shell). The $f_{total}(q)$ considers point scatterers in both the core and shell, the $f_{core}(q)$ only considers the point scatterers in the core, and the $f_{shell}(q)$ only considers the point scatterers in the shell.

To analyze experimental SANS profiles, we incorporate smearing effects by utilizing the scattering instrument's output variance $\sigma$ and the mean scattering vector $\bar{q}$.[58]

$$I_{X\,smeared}(q_i) = \int_0^\infty I_X(q) R(q, q_i)\, dq \tag{3}$$

$$R(q, \bar{q}) = \frac{1}{(2\pi\sigma)^{1/2}} exp\left(\frac{-(q-\bar{q})^2}{2\sigma}\right) \tag{4}$$

As was discussed previously,[52] when incorporating instrument smearing effects, one needs to calculate the I(q) at q values below and above the smallest and highest experimental q values. We note here that experimentally obtaining contrast matched SANS scattering profiles for all individual components can be quite laborious; we also evaluate how CREASE performs when only some of these (only total and core) scattering profiles are provided.

For calculating the computed scattering profiles, the Debye scattering approach is very computationally intensive as it must first evaluate the P(q) (Equation 2) and then combine that with the S(q) (Equation 1) to obtain the overall scattering intensity I(q). To overcome this issue, recent publications on CREASE describe implementation of a machine learning (ML) model (neural network) to directly relate the gene set to the corresponding scattering intensity profile.[39,52] For this work, we develop separate ML models linking the gene set to the S(q) and to the P(q). The rationale is to use the S(q) ML model more generally for other analogous concentrated primary particle solutions (*e.g.*, concentration solution of amphiphilic polymer vesicles) as the core-corona micelle/core-shell particle solutions. For the development of the ML models, we utilize a similar approach as our previous work with neural networks[52] and provide the complete details of the current implementations in **ESI Section S1**.

**II. C.  In silico systems tested/analyzed with 'P(q) and S(q) CREASE'**

We focus our *in silico* method validation on concentrated micelle solutions where we assume the micelle shell thickness scales with the core size (constant ratio of core radius to total radius). We consider all shell cases previously described when we apply this 'P(q) and S(q) CREASE' to the experimental scattering data at the end of this paper. After identifying and implementing the genes, the user then must set the limits (low and high value) for the genes. As was previously discussed,[52] the more information given to CREASE through the gene limits (*e.g.*, for known information, use smaller ranges of gene values to search over), the better the performance as CREASE can then limit the number of unknowns it must optimize for. For the *in silico* validation part of this work on computationally generated concentrated (generic core-shell) micelle solutions, we set the range for the micelle diameter as the true value (50 nm) ± 5 nm, the size dispersity as the true value (0.05 or 0.15) ± 0.025, the core-micelle ratio as 0.05 to 0.95, and micelle volume fraction as 0.05 to 0.5. We arbitrarily select $D_{micelle}$ to be 50 nm knowing that all steps in this method are agnostic to the micelle size and can be used to analyze systems with larger and smaller micelle sizes. For the *in silico* validation of this 'P(q) and S(q) CREASE', we consider a wide range of computationally generated concentrated micelle solution structures with different micelle size dispersity, ratio of core size to total micelle size, micelle volume fraction, and extent of micelle aggregation. The concentrated micelle solution 3D structures are generated by placing ~20,000 spheres in a large box. Each sphere is designated either as a micelle sphere (with a desired micelle size and a desired core to micelle diameter ratio selected from the distribution of dimensions) or a spacer sphere (*i.e.*, coarse representation of solvents) with diameter from a lognormal distribution with average of 50 nm and dispersity of 0.01. The relative number of micelles to spacer spheres is selected to achieve the desired micelle volume fraction (*e.g.*, for 0.40 volume fraction, we designate 8,000 of the 20,000 spheres as micelle spheres). Then the relative placement of micelle and spacer spheres is tuned to achieve a range of micelle aggregation states quantified by the varying values of the contact peak in the micelle center-to-center radial distribution function.

**ESI Table S1** provides the various parameters examined, and **ESI Figures S1-S12** provide the validation of 'P(q) and S(q) CREASE' on all 27 target systems. The target scattering profiles are produced by averaging 35 structures with similar characteristics to incorporate the inherent variability in the system like how a scattering experiment obtains scattering over a large sample volume. For the *in silico* systems, we compare the micelle information related to the P(q) (micelle diameter, micelle dispersity, core-micelle size ratio, and micelle concentration) and structural arrangement related to the S(q) (through the radial distribution function, RDF). For all systems, we perform three independent 'P(q) and S(q) CREASE' runs and compare the average and standard deviation from the three CREASE runs against the target structure. All visualizations are created using the VMD software.[59]

## II. D. Experiments and *in vitro* systems analyzed with 'P(q) and S(q) CREASE'

We consider penta(ethyleneglycol) monododecylether ($C_{12}E_5$) self-assembled on the surface hydrophilic silica nanoparticles (diameter ~ 30 nm) as an example *in vitro* core-shell particle system to test the applicability of CREASE. The $C_{12}E_5$ surfactant adsorbs onto the silica nanoparticle surface via weak H-bonds between ethoxylated headgroups of the surfactant molecule and silanol group on the nanoparticles. These H-bonds can either be direct between the headgroup and the surface or can be mediated with water molecules.[60] Because of the low adsorption free energy and bending rigidity of the surfactant bilayers, $C_{12}E_5$ is known to form discrete surface patches on the surface of highly curved silica nanoparticles.[61-63]

Here we investigate the change in the self-assembled structure of $C_{12}E_5$ coated silica nanoparticles upon increasing amounts of NaCl at 30°C and 40°C. We performed the SANS measurements at fixed concentration of silica nanoparticles (1 wt%) containing $C_{12}E_5$ at $4.5 \times 10^{-6}$ mol/m² of silica, which is equivalent to ~90% of the maximum surface excess of the surfactant at 20°C. The amount of surfactant adsorbed on the surface of silica nanoparticles increases with increasing temperature and salinity. Such aspects of adsorption will be discussed in a separate forthcoming publication. In the current study, nearly all of the added $C_{12}E_5$ adsorbs on the surface of silica nanoparticles and no free unadsorbed surfactant micelles exist in the solvent. The SANS studies were performed using $H_2O:D_2O$ mixture as solvents to generate two contrast scenarios: (A) *Shell contrast matched*: $H_2O:D_2O$ at 88:12 matches the SLD ($1.28 \times 10^{-5}$ nm$^{-2}$) of the surfactant shell thus the scattering solely originates from the silica core and provides direct information on the spatial distribution of the nanoparticles. (B) *Core contrast matched:* $H_2O:D_2O$ at 38:62 was used as a solvent to match the SLD ($3.5 \times 10^{-4}$ nm$^{-2}$) of silica core, enabling a selective determination of the changes in surfactant shell upon introducing NaCl and increasing temperature. The SANS experiments were performed at Oak Ridge National Laboratory (ORNL) SNS facility using the EQ-SANS instrument with pinhole collimation at a neutron wavelength of 6 Å. Further details about the SANS experiments can be found in previous publications.[64-66]

**III. Results and Discussion**

**III. A. Validation of 'P(q) and S(q) CREASE' on *in silico* concentrated solutions of core-shell type micelles**

We perform an expansive validation of the 'P(q) and S(q) CREASE' method against 27 different computationally generated concentrated micelle solution structures with varying micelle size dispersity, core-micelle ratios, micelle volume fraction, and degree of aggregation. The reader is directed to **ESI Section S2** that provides the performance of CREASE for all 27 systems. For brevity, in this main paper, we only present a select few systems to demonstrate how this 'P(q) and S(q) CREASE' method performs for varying degrees of aggregation at 'low' micelle volume fraction (**Figure 3**), varying core-micelle size ratio at 'low' micelle size dispersity (**Figure 4**), and varying core-micelle size ratio at 'high' micelle size dispersity (**Figure 5 and 6**).

In **Figure 3**, we apply the newly developed 'P(q) and S(q) CREASE' to a system of concentrated micelles with an average diameter of 50 nm, 5% lognormal size dispersity, core-micelle ratio of 0.25, and 15% micelle volume fraction with varying degree of aggregation. **Figure 3a** provides the scattering profile comparisons with the black line as the target scattering profiles input into the 'P(q) and S(q) CREASE', the red line as the computed scattering profile from the output of 'P(q) and S(q) CREASE' when all three target scattering profiles ($I_{micelle}(q)$, $I_{shell}(q)$, and $I_{core}(q)$) are input, and the blue line as the computed scattering profile from the output of 'P(q) and S(q) CREASE' when only two target scattering profiles ($I_{micelle}(q)$ and $I_{core}(q)$) are input. The final case of 'P(q) and S(q) CREASE' illustrates how CREASE performs with inputs from fewer performed scattering experiments, potentially saving the user small angle scattering beam time. While $I_{shell}(q)$ is not input for the final case, we compute the $I_{shell}(q)$ from the best structure returned by CREASE to illustrate the close scattering match to the cases when that $I_{shell}(q)$ is input.

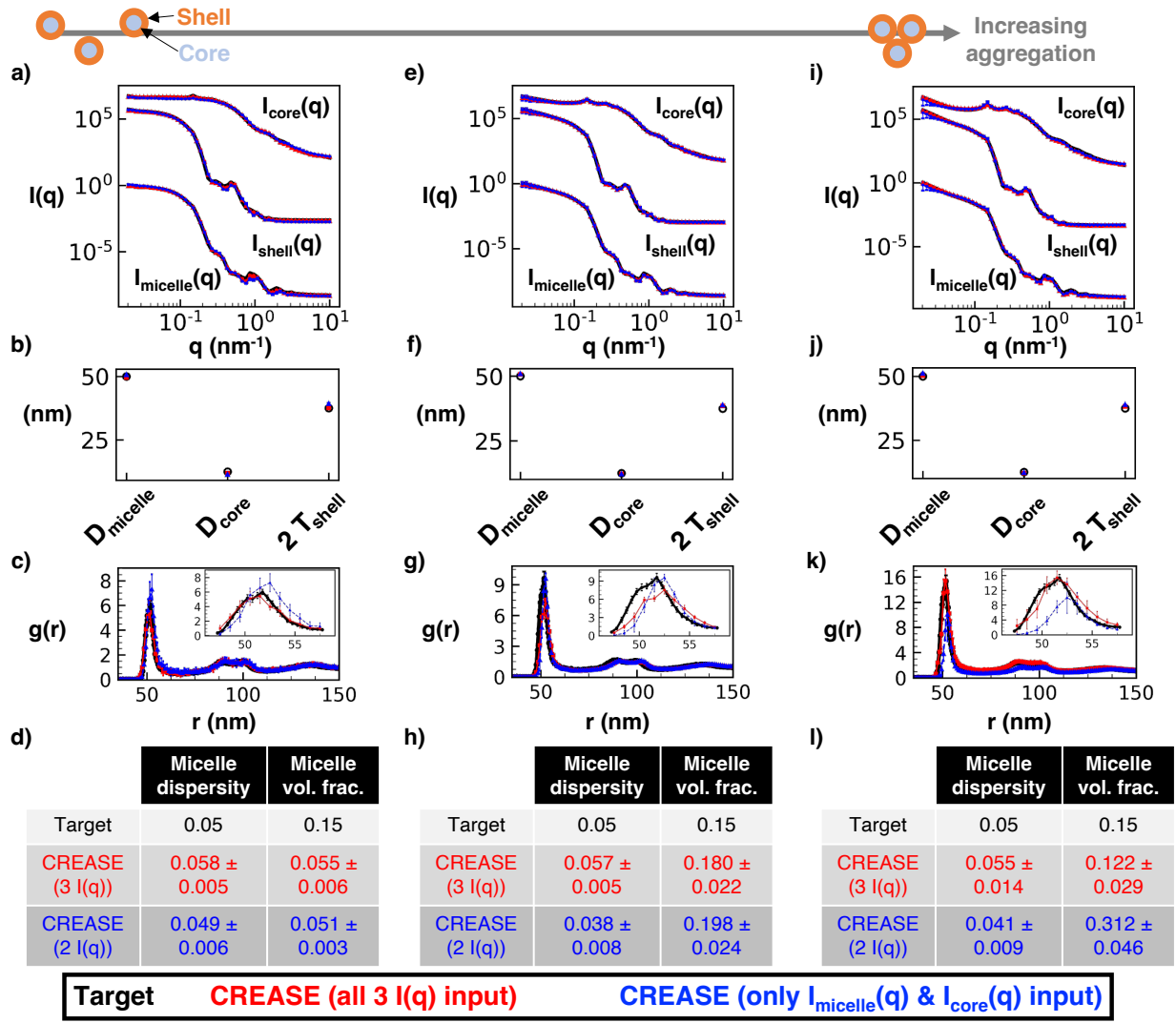

***Figure 3***: *'P(q) and S(q) CREASE' applied to a concentrated micelle solution with a 50 nm average diameter, 0.05 micelle size dispersity, 0.25 core-micelle size ratio, and 0.15 micelle volume fraction to simultaneously identify the information about the micelles and the structural arrangement. a) Scattering intensity, I(q), for the target structure (black) with disperse micelle arrangement, CREASE with all three I(q) used as inputs (red), and CREASE with only the $I_{micelle}(q)$ and $I_{core}(q)$ used as inputs (blue). While the blue case only receives two I(q) curves as inputs, we calculate the $I_{shell}(q)$ from the output structure for comparison. b) Micelle diameter ($D_{micelle}$), core diameter ($D_{core}$), and twice the shell thickness (2 $T_{shell}$) for the target and CREASE outputs. c)*

*Micelle structural arrangement is quantified using the micelle-micelle radial distribution function (RDF) comparing the target and CREASE outputs. d) Additional information on the micelle size dispersity and micelle volume fraction that the target possesses, and the CREASE methods converge to. e-h) Are the same as a-d) with increasing aggregation to a weakly aggregating target system. i-l) Are the same as a-d) with increasing aggregation to a strongly aggregating target system. The error bars are the standard deviation of the average of 3 independent runs of the 'P(q) and S(q) CREASE' method.*

**Figure 3b** and **Figure 3c** display information on the micelle dimensions returned by CREASE compared to the set values in the target structure (related to P(q)) and the inter-micelle structure held by the radial distribution function, RDF, (related to S(q)). As can be seen in **Figure 3b**, both 'P(q) and S(q) CREASE' versions, one with all three scattering profiles input and one with only two scattering profiles input, obtain similar micelle dimensions to the target micelle values. The micelle-micelle RDF shows a quantitative match between the target and the 'P(q) and S(q) CREASE' with all three scattering curves input (red) and a close match with the other CREASE version with only two scattering curves input (blue). **Figure 3d** provides additional comparisons between the CREASE runs and the target values for micelle size dispersity and micelle volume fraction. While both 'P(q) and S(q) CREASE' versions achieve quantitative matches for the micelle size dispersity, CREASE underpredicts the micelle volume fraction. We expect CREASE to poorly predict micelle volume fraction at conditions (either low concentration or insignificant particle aggregation at high concentration) where the effect of volume fraction is minimal. As we consider systems with increasing micelle aggregation, **Figures 3e-h** and **Figures 3i-l**, both 'P(q) and S(q) CREASE' versions achieve close scattering matches and micelle size dimensions (**Figures 3e-f, 3i-j**). For the intermediate degree of micelle aggregation in **Figure 3g**, the 'P(q) and S(q) CREASE' with only two scattering profile inputs achieves a quantitative match while the 'P(q) and S(q) CREASE' with all three scattering profiles input predicts a structure with lower RDF peak (less aggregation) than the target structure. Both 'P(q) and S(q) CREASE' versions achieve similarly close predictions for the micelle size dispersity and micelle volume fraction in **Figure 3h**. For the highest degree of micelle aggregation in **Figure 3k**, CREASE with all three scattering inputs achieves a better quantitative match than CREASE with only two scattering inputs. The 'P(q) and S(q) CREASE' with only two scattering profiles input converges to a structure with an RDF lower than the target value; however, this mismatch is caused by the substantial overprediction of the micelle volume fraction in **Figure 3l** that causes the RDF to achieve a lower peak value. In **Figure 3l**, we observe the 'P(q) and S(q) CREASE' with all three scattering profile inputs to achieve a quantitative match for both micelle size dispersity and micelle volume fraction while the 'P(q) and S(q) CREASE' with two scattering profile inputs only achieves a quantitative match for the micelle size dispersity.

In contrast to **Figure 3** that focuses on the effect of aggregation, **Figure 4** highlights how changing the core-micelle size ratio affects the performance of the 'P(q) and S(q) CREASE.' In **Figure 4** 'P(q) and S(q) CREASE' is applied to a system of concentrated micelles with an average diameter of 50 nm, 5% lognormal size dispersity, 40% micelle volume fraction, an intermediate degree of micelle aggregation, and various core-micelle ratios. We find that both 'P(q) and S(q) CREASE' approaches can reconstruct concentrated micelle solutions with quantitatively similar scattering profiles and micelle dimensions regardless of the core-micelle ratio. The extent of match in micelle-micelle RDF between the target structure and the 'P(q) and S(q) CREASE' methods, are on average relatively precise with minor differences in the peak RDF value. Interestingly, we observe that the 'P(q) and S(q) CREASE' method that only takes two I(q) inputs (blue) on average achieves closer matches between its predicted RDF and the target RDF compared to the 'P(q) and S(q) CREASE' with all three I(q) profiles input. This finding is counterintuitive as one would expect the incorporation of additional information (an extra I(q) profile) to result in an improved structure reconstruction. However, the 'P(q) and S(q) CREASE' utilizes an ML model to predict the computed scattering profiles, and the $I_{shell}(q)$ profile tends to possess numerous sharp features (oscillations) that reduce the ML model performance at predicting the exact values for each peak and trough. Thus, the 'P(q) and S(q) CREASE' that only utilizes the $I_{micelle}(q)$ and $I_{core}(q)$ that have fewer (and less sharp) features than $I_{shell}(q)$, performs slightly better. When we examine the predicted micelle size dispersity and micelle volume fraction from both 'P(q) and S(q) CREASE' versions, both CREASE methods achieve close predictions to the target values.

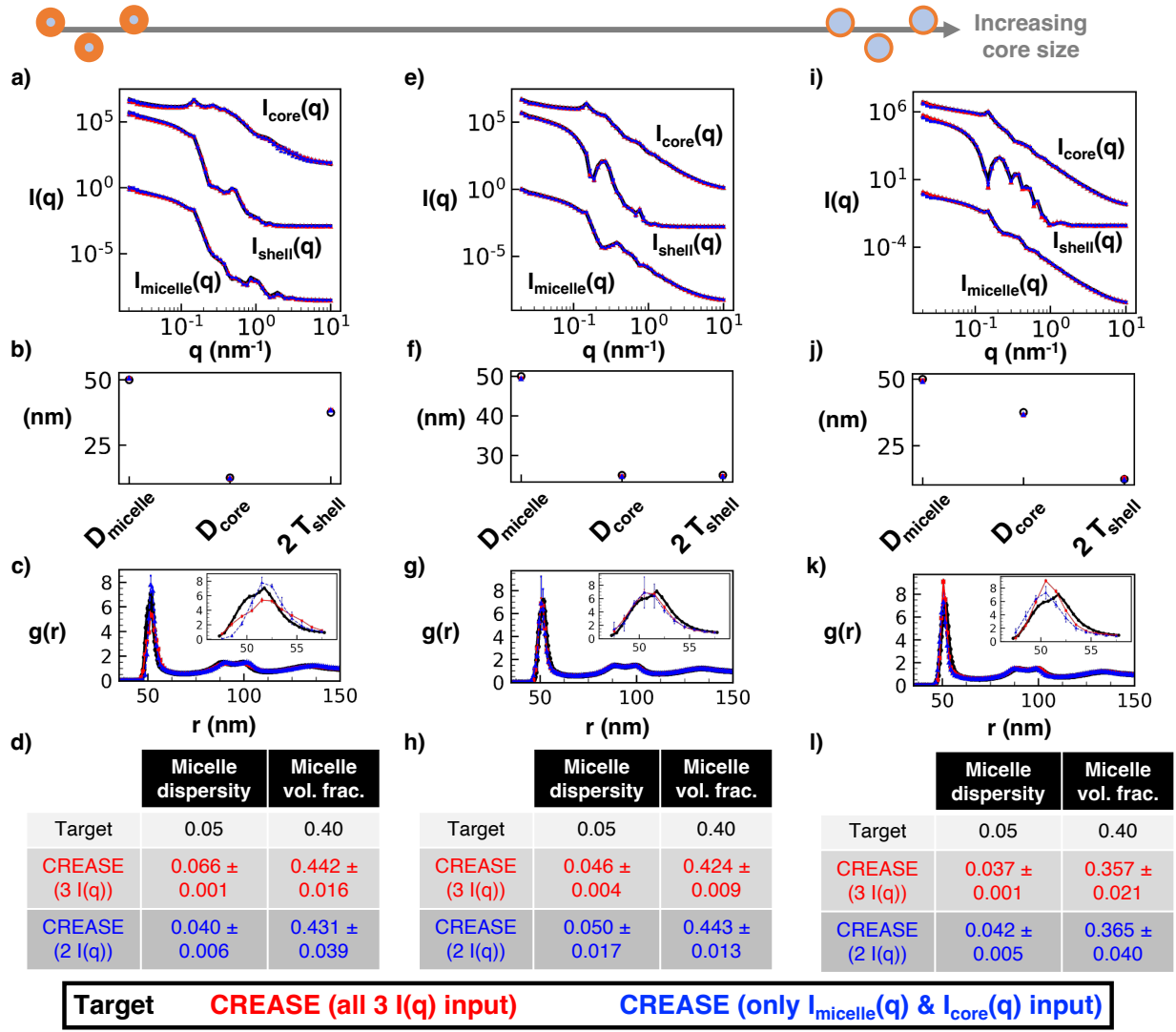

*Figure 4*: *'P(q) and S(q) CREASE' applied to a concentrated micelle solution with a 50 nm average diameter, 0.05 micelle size dispersity, 0.40 micelle volume fraction, and intermediate micelle aggregation to simultaneously identify the information about the micelle and the structural arrangement. a) Scattering intensity, I(q), for the target structure (black) with disperse micelle arrangement, CREASE with all three I(q) used as inputs (red), and CREASE with only the $I_{micelle}(q)$ and $I_{core}(q)$ used as inputs (blue). While the blue case only receives two I(q) curves as inputs, we calculate the $I_{shell}(q)$ from the output structure for comparison. b) Micelle diameter ($D_{micelle}$), core diameter ($D_{core}$), and twice shell thickness (2 $T_{shell}$) for the target and CREASE outputs. c) Micelle*

*structural arrangement is quantified using the micelle-micelle radial distribution function (RDF) comparing the target and CREASE outputs. d) Additional information on the micelle size dispersity and micelle volume fraction that the target possesses, and the CREASE methods converge to. a-d) Are for a system with 0.25 core-micelle size ratio. e-h) Are for a system with 0.50 core-micelle size ratio. i-l) Are for a system with 0.75 core-micelle size ratio. The error bars are the standard deviation of the average of 3 independent runs of the 'P(q) and S(q) CREASE' method.*

Next in **Figure 5**, we consider the performance of 'P(q) and S(q) CREASE' on a system of concentrated micelles with diameter, volume fraction, extent of aggregation same as **Figure 4** at higher dispersity - 15% lognormal size dispersity in micelle diameter. We find that, unsurprisingly, the increased micelle size dispersity reduces the match between the 'P(q) and S(q) CREASE' results and the target system because the increased micelle size dispersity results in smoother scattering profiles with fewer features (*e.g.*, reader can compare scattering profiles in **Figure 5a, e, i** to those in **Figure 4a, e, i**). At higher micelle size dispersity, there are more degenerate solutions (different P(q) and S(q) combinations that give similar scattering profiles) resulting in 'P(q) and S(q) CREASE' outputs that are farther from the target values. One can find close scattering matches between the target and 'P(q) and S(q) CREASE' methods, yet both 'P(q) and S(q) CREASE' variations erroneously converge to a smaller micelle diameter and lower micelle size dispersity than that of the target system. As a result, the RDF plots demonstrate substantial deviations from the target and 'P(q) and S(q) CREASE' outputs because the smaller dimeter and dispersity result in the primary peak shift to the left and a narrowing of the peak (causing a higher peak value). We find these trends to be consistent regardless of the ratio of the core size to total micelle size and extent of micelle aggregation. For systems with smoother scattering profiles with fewer features (*e.g.*, those with high micelle size dispersity), the output from the 'P(q) and S(q) CREASE' method may shift from quantitative to qualitative reconstructions if the user forces 'P(q) and S(q) CREASE' to converge for all details about the system without additional information to assist with eliminating degenerate solutions. If the user inputs additional relevant information about the system (obtained from other characterization techniques like microscopy), 'P(q) and S(q) CREASE' can avoid those degenerate solutions to converge to the correct P(q) and S(q). In **Figure 6**, we illustrate how providing additional information about the micelle size and size dispersity for the same systems as **Figure 5**, leads to a substantially more quantitative match between the target and output from 'P(q) and S(q) CREASE' than that seen in **Figure 5.**

*Figure 5: Same as **Figure 4** with a 50 nm average diameter, **0.15 micelle size dispersity**, 0.40 micelle volume fraction, and intermediate micelle aggregation.*

In **Figure 6**, we demonstrate the performance of 'P(q) and S(q) CREASE' on a system of concentrated micelles with an average diameter of 50 nm, 15% lognormal size dispersity, 40% micelle volume fraction, an intermediate degree of micelle aggregation, and various core-micelle ratios (same system as **Figure 5**) when 'P(q) and S(q) CREASE' is provided a *smaller* range of micelle diameter and micelle size dispersity than is provided to obtain **Figure 5** results. This is to demonstrate the case when the user has additional information about micelle size and dispersity from microscopy or other characterization (*e.g.*, cryo-transmission electron microscopy[35]) and how the inclusion of addition information into 'P(q) and S(q) CREASE' improves its performance at high micelle size dispersity. For both 'P(q) and S(q) CREASE', we input the micelle diameter range as the target value (50 nm) ± 1 nm and the micelle size dispersity as the target value (0.15) ± 0.01. We highlight this difference in inputs by plotting the 'P(q) and S(q) CREASE' with all three I(q) inputs in orange and the 'P(q) and S(q) CREASE' with only two I(q) inputs in cyan. The scattering profile matches between the target and both 'P(q) and S(q) CREASE' in **Figure 6** are quantitatively similar as in **Figure 5**. However, by including additional information about the micelle total size and size dispersity, both 'P(q) and S(q) CREASE' methods converge to quantitatively similar values for the micelle core and shell sizes. The RDF profiles from the target system and from these 'P(q) and S(q) CREASE' methods are a nearly perfect match. Interestingly, we find that both 'P(q) and S(q) CREASE' converge to higher micelle volume fractions than the target systems. We note that a user could also extract information about the micelle volume fraction during the proposed characterization to input smaller ranges for the total micelle diameter and size dispersity. Overall, **Figure 6** demonstrates that 'P(q) and S(q) CREASE' can still provide quantitatively similar P(q) and S(q) reconstructions for systems with relatively feature-less scattering profiles if the user can provide additional information that can be used to reduce the range of parameters the CREASE method has to optimize over.

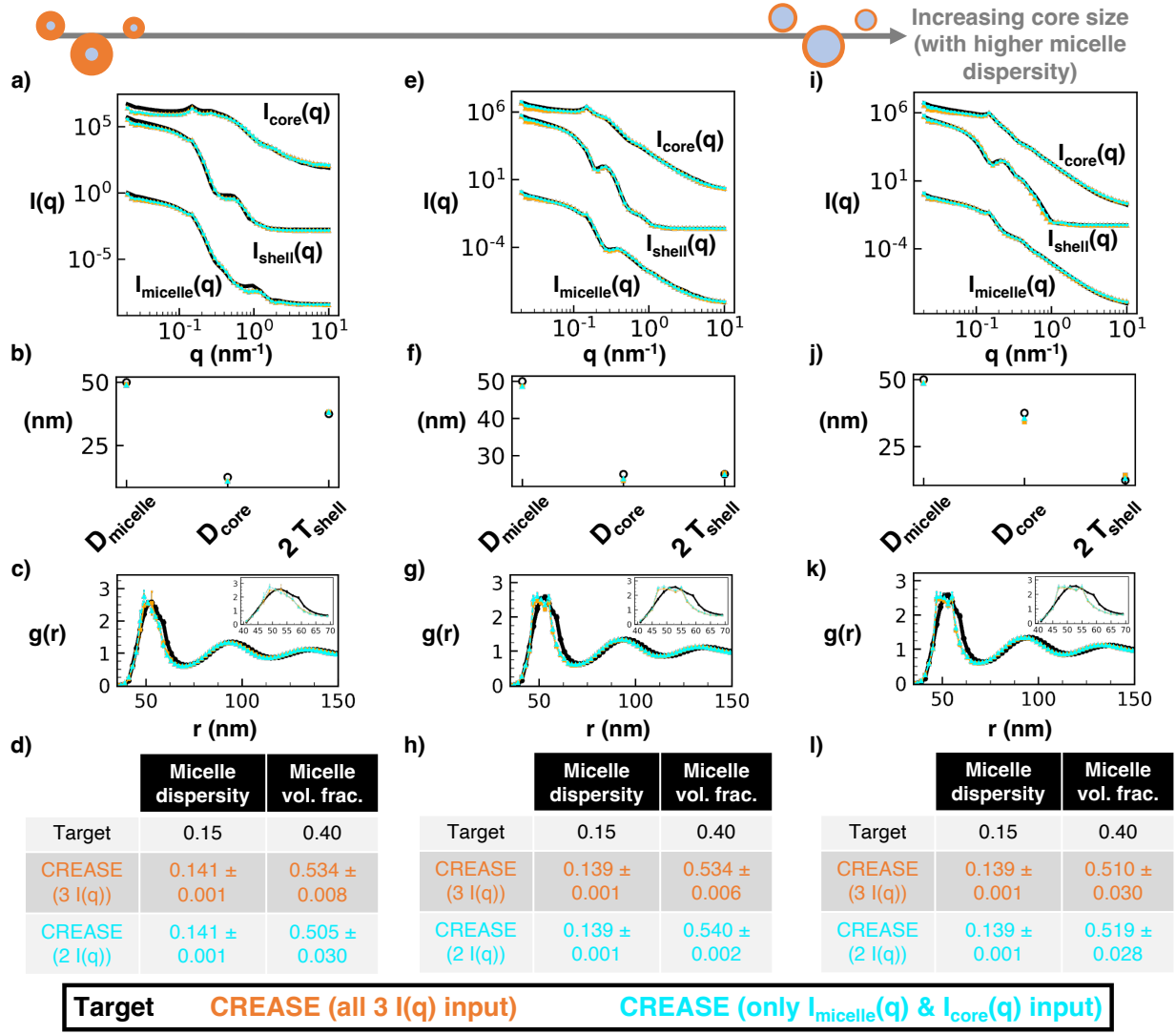

*Figure 6: Same as **Figure 4** with a 50 nm average diameter, **0.15 micelle size dispersity,** 0.40 micelle volume fraction, and intermediate micelle aggregation. For this Figure, the 'P(q) and S(q) CREASE' is provided a smaller range of micelle diameter and micelle size dispersity to demonstrate how the inclusion of addition information into 'P(q) and S(q) CREASE' improves its performance at high micelle size dispersity. Experimentally, one could perform cryo-TEM imaging to obtain an approximate micelle diameter and size dispersity. For both 'P(q) and S(q) CREASE', we set the micelle diameter as the target value (50 nm) ± 1 nm and the micelle size dispersity as the target value (0.15) ± 0.01. We highlight this difference in inputs by plotting the 'P(q) and S(q)*

*CREASE' with all three I(q) inputs in orange and the 'P(q) and S(q) CREASE' with only two I(q) inputs in cyan.*

**III. B. Application of 'P(q) and S(q) CREASE' to nanoparticles with adsorbed surfactant molecules**

Having validated the 'P(q) and S(q) CREASE' approach on *in silico* systems of concentrated core-shell micelles in **Section III.A.**, in this section, we apply the method to analyze scattering results from a system of core-shell nanoparticles, specifically nonionic amphiphilic n-alkyl pentaethylene glycol monododecyl ether ($C_{12}E_5$) surfactants coated silica nanoparticles. We consider a range of conditions (salt concentration and temperature); no salt and low temperature produce a disperse solution and increasing salt concentration and temperature leads to increasing nanoparticle aggregation. The nanoparticle core has a known diameter of 30 nm and lognormal size dispersity of 0.1. The structural features of the surfactant shell are not known and there is interest in understanding how it varies with changing salt concentration and temperature. To obtain that understanding, we adapt the 'P(q) and S(q) CREASE' method to consider several scenarios to test hypotheses about how (if) the surfactant shell varies when adsorbed to the polydisperse nanoparticles. These hypotheses as described in the Methods Section in detail. Briefly, we adapt 'P(q) and S(q) CREASE' to consider the following scenarios (colors denoted in parenthesis serve as legend for figures and tables describing these results):

1) constant surfactant shell thickness with no overlap possible between neighboring coated nanoparticles (red color)

2) surfactant shell thickness scales with the nanoparticle size due to experiments finding polydispersity in nanoparticle diameter; no overlap possible between neighboring coated nanoparticles (blue color)

3) surfactant shell has an average thickness and thickness dispersity which are independent of the nanoparticle's size and size dispersity; no overlap possible between neighboring coated nanoparticles (purple color)

4) constant surfactant shell thickness with some overlap possible between neighboring coated nanoparticles (orange color)

5) surfactant shell thickness scales with the nanoparticle size due to polydispersity in nanoparticle size with some overlap possible between neighboring coated nanoparticles (cyan color)

6) surfactant shell has an average thickness and thickness dispersity which are independent of the nanoparticle size and allow some overlap between neighboring coated nanoparticles (magenta color)

**Figure 7** provides the comparisons between the input experimental nanoparticle and surfactant shell contrast matched scattering profiles and results from the above six 'P(q) and S(q) CREASE' variations for three values of salt concentration and two temperatures. For clearer (less crowded) view of the data in **Figure 7, ESI Figure S13-S15** provides those same scattering profiles shown in **Figure 7** but split into the cases that allow 'primary particle' (core-shell nanoparticle) overlap or not. To quantify the closeness between the 'P(q) and S(q) CREASE' output scattering profiles and the experimental scattering profiles, we calculate the $\chi^2$ value with a smaller value indicating a closer match between CREASE and experimental scattering profile.[52]

For the less aggregated systems (0 mM at both temperatures 30°C or 45°C) **(Figure 7a and 7b)** we find that the performance of 'P(q) and S(q) CREASE' variations (1 – 3) that do not allow overlap is better than that of 'P(q) and S(q) CREASE' variations (4 – 6) that allow for overlap. However, with increasing extents of aggregation, the trend is flipped, and the 'P(q) and S(q) CREASE' variations (4 – 6) that allow for core-shell particle overlap show lower $\chi^2$ values (better fits) on average than 'P(q) and S(q) CREASE' variations (1 – 3) that do not allow overlap. This finding suggests that these surfactant-coated nanoparticles are not hard core-shell particles but that upon aggregation these core-shell nanoparticles may share common shell regions. Within the specific variations of 'P(q) and S(q) CREASE', we find that case 5 (cyan) consistently achieves the lowest/lower $\chi^2$ value for all salt concentrations and temperatures considered. Thus, those results suggest a potential scenario where the surfactant shell thickness likely scales with the nanoparticle size with smaller (larger) nanoparticles possessing thinner (thicker) surfactant shells. However, it is not clear as to why or how that would occur physically in these surfactant-coated nanoparticle systems, motivating a direction for future study.

**Table 1** provides the surfactant shell dimensions for all solution conditions and 'P(q) and S(q) CREASE' variations we consider in **Figure 7**. For both solution temperatures considered, we consistently find that in most CREASE variations the shell thickness increases as we go from 0mM to 2mM salt concentration and then decreases from 2mM to 5mM salt concentration. Furthermore, the 5mM salt concentration for both temperatures considered has smallest shell thickness among all variations. When we consider the cases that allow overlap between the core-shell particles, we find that the shell overlap increases with salt concentration from 0mM to 5mM. When we consider the solution temperature effect, we find that the higher temperature (45°C) decreases the surfactant shell thickness for all salt concentrations (0mM, 2mM, 5mM) for all cases considered except Case 4 (orange) which results in a minor increase in surfactant shell thickness at higher temperature. Overall, **Table 1** suggests that both solution temperature and salinity significantly affect the extent of 'primary particle' aggregation and the extent of aggregation dictates the surfactant shell structure – size and extent of overlap.

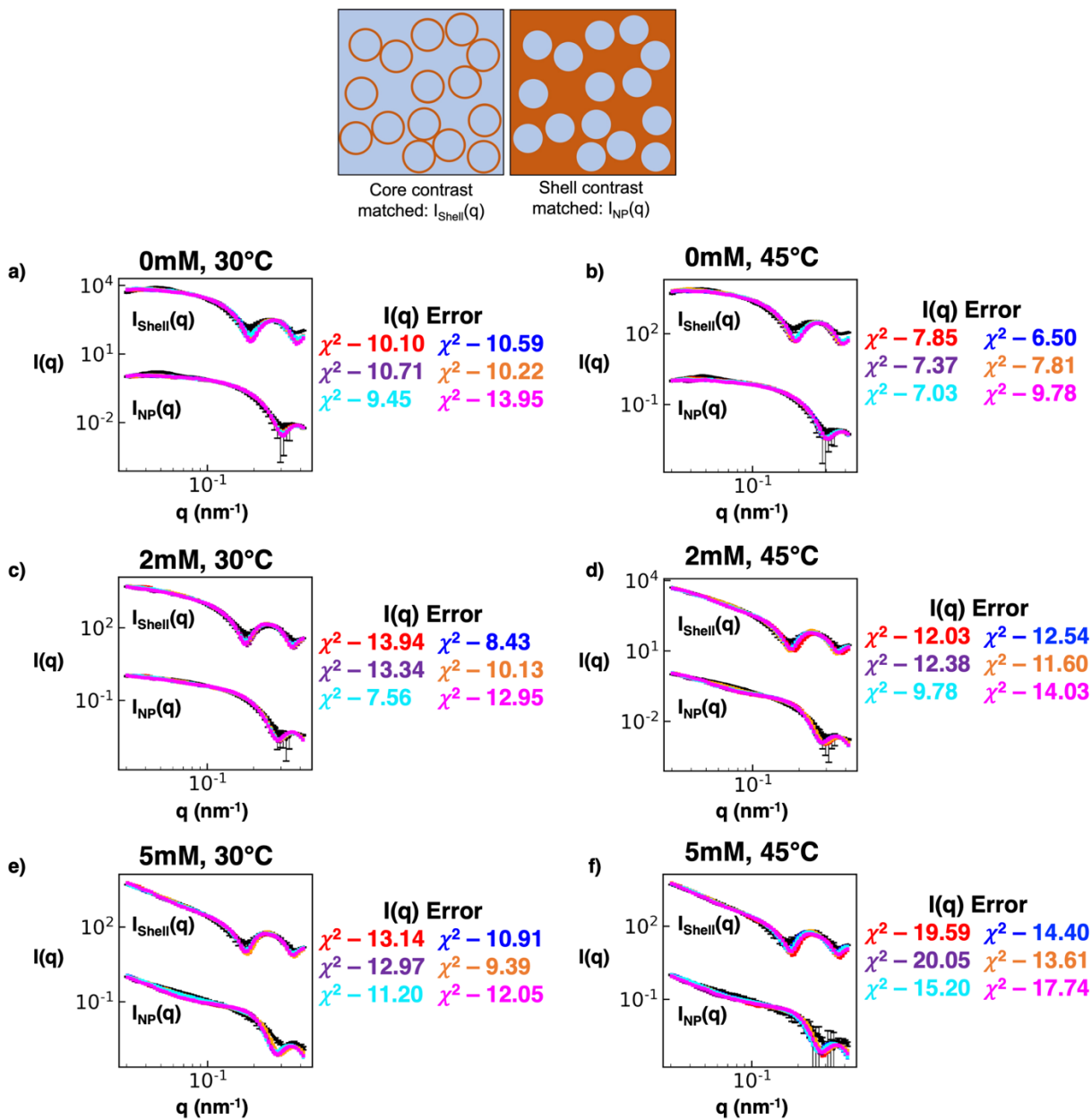

*Figure 7*: *'P(q) and S(q) CREASE' applied to experimental small angle neutron scattering of surfactant adsorbed nanoparticles to simultaneously identify the nanoparticle and surfactant shell dimensions and the structural arrangement in solution for various solution temperatures and salt concentrations. a-f) Scattering intensity, I(q), from the surfactant shell, $I_{shell}(q)$, and from the nanoparticle, $I_{NP}(q)$. The black line is the experimental contrast matched small angle neutron scattering profile, and the colored lines are the scattering profile results from various versions of 'P(q) and S(q) CREASE'. The red line is 'P(q) and S(q) CREASE' assuming a constant surfactant shell thickness; the blue line is 'P(q) and S(q) CREASE' assuming the surfactant shell thickness scales with the nanoparticle size; the purple line is 'P(q) and S(q) CREASE' assuming the surfactant shell has an average thickness and thickness dispersity independent of the nanoparticle; the orange line is 'P(q) and S(q) CREASE' assuming a constant surfactant shell thickness and allowing overlap between neighboring coated nanoparticles; the cyan line is 'P(q) and S(q) CREASE' assuming the surfactant shell thickness scales with the nanoparticle size and allowing overlap between neighboring coated nanoparticles; the magenta line is 'P(q) and S(q) CREASE' assuming the surfactant shell has an average thickness and thickness dispersity independent of the nanoparticle and allowing overlap between neighboring coated nanoparticles. a) and b) are the surfactant coated nanoparticle scattering at 0mM salt concentration and 30°C or 45°C, respectively. c) and d) are the surfactant coated nanoparticle scattering at 2mM salt concentration and 30°C or 45°C, respectively. e) and f) are the surfactant coated nanoparticle scattering at 5mM salt concentration and 30°C or 45°C, respectively. The $\chi^2$ value is a quantitative measure of the scattering matches between the target (black) and 'P(q) and S(q) CREASE' variation with a lower value indicating a closer fit. The error bars are the experimental SANS standard deviation and the standard deviation of the average of 3 independent runs of the 'P(q) and S(q) CREASE'.*

*Table 1*: Surfactant shell dimensions for the six versions of 'P(q) and S(q) CREASE' variations shown in **Figure 7**.

| CREASE variations | 0mM, 30°C | | 0mM, 45°C | |
|---|---|---|---|---|
| **Constant shell thickness** | **Thickness:** | 3.35 ± 0.12 nm | **Thickness:** | 2.73 ± 0.41 nm |
| **Shell scales with NP size** | **Thickness:** | 4.81 ± 1.03 nm | **Thickness:** | 4.36 ± 1.05 nm |
| | **Shell size ratio:** | 14.1 ± 3.0 % | **Shell size ratio:** | 13.1 ± 3.1 % |
| **Shell has average & dispersity** | **Ave. thickness:** | 3.67 ± 1.15 nm | **Ave. thickness:** | 3.57 ± 1.31 nm |
| | **Dispersity:** | 3.41 ± 0.63 % | **Dispersity:** | 6.74 ± 0.95 % |
| **Constant shell thickness w/ overlap** | **Thickness:** | 3.64 ± 0.51 nm | **Thickness:** | 3.73 ± 0.45 nm |
| | **Overlap:** | 2.42 ± 1.74 % | **Overlap:** | 1.72 ± 1.02 % |
| **Shell scales with NP size w/ overlap** | **Thickness:** | 5.97 ± 2.16 nm | **Thickness:** | 4.39 ± 2.03 nm |
| | **Shell size ratio:** | 17.0 ± 6.2 % | **Shell size ratio:** | 12.9 ± 5.9 % |
| | **Overlap:** | 2.76 ± 1.31 % | **Overlap:** | 5.95 ± 3.01 % |
| **Shell has average & dispersity w/ overlap** | **Ave. thickness:** | 3.16 ± 0.46 nm | **Ave. thickness:** | 2.97 ± 0.24 nm |
| | **Dispersity:** | 5.45 ± 3.19 % | **Dispersity:** | 7.98 ± 1.37 % |
| | **Overlap:** | 1.13 ± 0.70 % | **Overlap:** | 4.69 ± 2.53 % |
| **CREASE variations** | **2mM, 30°C** | | **2mM, 45°C** | |
| **Constant shell thickness** | **Thickness:** | 4.02 ± 0.07 nm | **Thickness:** | 3.19 ± 0.46 nm |
| **Shell scales with NP size** | **Thickness:** | 5.98 ± 0.45 nm | **Thickness:** | 5.66 ± 1.03 nm |
| | **Shell size ratio:** | 16.3 ± 1.2 % | **Shell size ratio:** | 15.1 ± 2.7 % |
| **Shell has average & dispersity** | **Ave. thickness:** | 3.10 ± 0.35 nm | **Ave. thickness:** | 3.02 ± 0.36 nm |
| | **Dispersity:** | 3.65 ± 1.57 % | **Dispersity:** | 6.55 ± 2.95 % |
| **Constant shell thickness w/ overlap** | **Thickness:** | 3.97 ± 0.58 nm | **Thickness:** | 4.27 ± 0.21 nm |
| | **Overlap:** | 3.58 ± 1.56 % | **Overlap:** | 5.82 ± 2.26 % |
| **Shell scales with NP size w/ overlap** | **Thickness:** | 6.73 ± 0.17 nm | **Thickness:** | 5.29 ± 0.64 nm |
| | **Shell size ratio:** | 17.9 ± 0.4 % | **Shell size ratio:** | 14.3 ± 1.7 % |
| | **Overlap:** | 2.83 ± 0.99 % | **Overlap:** | 3.81 ± 0.91 % |
| **Shell has average & dispersity w/ overlap** | **Ave. thickness:** | 3.66 ± 0.35 nm | **Ave. thickness:** | 2.79 ± 0.26 nm |
| | **Dispersity:** | 6.48 ± 2.71 % | **Dispersity:** | 3.95 ± 1.98 % |
| | **Overlap:** | 5.21 ± 2.30 % | **Overlap:** | 4.20 ± 1.28 % |
| **CREASE variations** | **5mM, 30°C** | | **5mM, 45°C** | |
| **Constant shell thickness** | **Thickness:** | 3.24 ± 0.21 nm | **Thickness:** | 2.38 ± 0.33 nm |
| **Shell scales with NP size** | **Thickness:** | 2.44 ± 0.16 nm | **Thickness:** | 2.28 ± 0.45 nm |
| | **Shell size ratio:** | 7.0 ± 0.5 % | **Shell size ratio:** | 6.4 ± 1.3 % |
| **Shell has average & dispersity** | **Ave. thickness:** | 2.79 ± 0.16 nm | **Ave. thickness:** | 3.10 ± 0.09 nm |
| | **Dispersity:** | 6.53 ± 2.31 % | **Dispersity:** | 5.48 ± 2.72 % |
| **Constant shell thickness w/ overlap** | **Thickness:** | 3.31 ± 0.19 nm | **Thickness:** | 3.52 ± 0.41 nm |
| | **Overlap:** | 5.49 ± 1.51 % | **Overlap:** | 8.99 ± 0.66 % |
| **Shell scales with NP size w/ overlap** | **Thickness:** | 2.65 ± 0.32 nm | **Thickness:** | 2.67 ± 0.51 nm |
| | **Shell size ratio:** | 7.6 ± 0.9 % | **Shell size ratio:** | 7.4 ± 1.4 % |
| | **Overlap:** | 6.57 ± 2.15 % | **Overlap:** | 6.00 ± 1.91 % |
| **Shell has average & dispersity w/ overlap** | **Ave. thickness:** | 3.04 ± 0.50 nm | **Ave. thickness:** | 2.67 ± 0.04 nm |
| | **Dispersity:** | 3.41 ± 1.17 % | **Dispersity:** | 3.23 ± 2.33 % |
| | **Overlap:** | 6.00 ± 1.22 % | **Overlap:** | 7.81 ± 2.62 % |

## IV. Conclusion

In this paper we presented an open-source 'P(q) and S(q) CREASE' computational method to interpret the form and spatial arrangement of primary particles in concentrated solutions from small angle scattering experiments; primary particles of interest are core-shell micelles or core-shell surfactant coated nanoparticles. We validated this combined form factor, P(q), and structure factor, S(q), CREASE approach on *in silico* concentrated micelle solutions with various micelle size dispersity, micelle concentration, core-micelle ratios, and degrees of micelle aggregation. For every system considered, we quantified the performance of this 'P(q) and S(q) CREASE' between its output and the target structure by comparing the error between the target and CREASE scattering profiles, converged micelle dimensions, and radial distribution functions. Additionally, we incorporated machine learning (ML) in a compartmentalized manner by independently developing an artificial neural network for the system's form factor P(q) and structure factor S(q), enabling rapid transferability to other related systems such as solutions of concentrated vesicles by simply adjusting the ML model for the form factor. We concluded this work by applying this 'P(q) and S(q) CREASE' method to surfactant adsorbed core-shell nanoparticles where the surfactant shell varies based on the solutions' salt concentration and temperature. We explored how various assumptions about the surfactant shell thickness resulted in corresponding differences in the scattering profile match between 'P(q) and S(q) CREASE' and the experimental data and identified the assumption that consistently provided the closest scattering match as the most likely explanation for the surfactant shell structure.

**Electronic Supporting Information (ESI):**

- Details of the machine learning models (ANNs) utilized in this study. 'P(q) and S(q) CREASE' performance on all *in silico* concentrated micelle solution cases considered. 'P(q) and S(q) CREASE' applied to surfactant coated nanoparticles.

**Data availability**:

All data is available for research use upon request from the corresponding author.

**Acknowledgements**:

C.M.H. and A.J. acknowledge financial support from the Air Force Office of Scientific Research (MURI-FA 9550-18-1-0142). B.B. and Y.M. thank Dr. Gergely Nagy and Dr. William Heller for their assistance with the SANS data acquisition. B.B. and Y.M. acknowledge the financial support by ACS – Petroleum Research Funds. This research used resources at the Spallation Neutron Source, DOE Office of Science User Facilities operated by the Oak Ridge National Laboratory. This work was supported with computational resources from the University of Delaware (Caviness cluster).

**Author Contributions:**

C.M.H. was responsible for programming, computing, and data analysis under the advisement of A.J. Y.M. and B.B. measured and provided the SANS data of the surfactant coated nanoparticles. C.M.H. and A.J. wrote the manuscript. C.M.H., B.B., and A.J. edited the manuscript.

**Competing Interests:**

The authors declare no competing interests.